# 1305 nm MoTe$_2$-on-silicon Laser

Hanlin Fang[1,2], Jin Liu[2], Hongji Li[1,2], Lidan Zhou[1], Lin Liu[1], Juntao Li[1,2*], Xuehua Wang[1,2], Thomas F. Krauss[3], Yue Wang[3]

[1] State Key Laboratory of Optoelectronic Materials and Technologies, Sun Yat-Sen University, Guangzhou, 510275, China

[2] School of Physics, Sun Yat-Sen University, Guangzhou, 510275, China

[3] Department of Physics, University of York, York, YO10 5DD, UK

*correspondence to: lijt3@mail.sysu.edu.cn

**ABSTRACT**

The missing piece in the jigsaw of silicon photonics is a light source that can be easily incorporated into the standard silicon fabrication process. Recent advances in the development of atomically thin layers of semiconducting transition metal dichalogenides (TMDs), with direct bandgaps in the near-infrared region, have opened up new possibilities for addressing this need. Here, we report a unique silicon laser source that employs molybdenum ditelluride (MoTe$_2$) as a gain material in a photonic crystal nanocavity resonator, fabricated in silicon-on-insulator. We demonstrate optically pumped MoTe$_2$-on-silicon devices lasing at 1305 nm, i.e. in the centre of the O-band used in optical communications, operating in the continuous-wave (CW) regime, at room temperature and with a threshold power density as low as 1.5 kW/cm$^2$. This 2D-on-silicon geometry offers the promise of an integrated low-cost electrically pumped nanoscale silicon light source, thereby adding an essential building block to the silicon photonics platform.

# INTRODUCTION

Due to its excellent integratability and compatibility with CMOS technology, silicon photonics has been widely recognized as the most promising platform a for future broadband, high-speed data transmission infrastructure [1]. Photonics offers substantial reductions in operating energy and improvements in the performance of data communication systems, which for instance explains its wide use in datacentres [2]. While many essential silicon photonics components have already been demonstrated with excellent performance, such as extremely low-loss waveguides [3], ultra-fast modulators [4], and high bandwidth detectors [5], a silicon light source that can be easily incorporated into the standard fabrication process has remained challenging. Silicon itself is a poor light emitter due to its indirect bandgap, so a combination with suitable light emitting materials that can be directly incorporated into the process flow needs to be investigated. Most notably, the hybrid integration of III-V materials with silicon has made significant progress [6,7], but the approach is costly and complex, so there is still an appetite for exploring alternative solutions. The $MoTe_2$-on-silicon laser we present here is such a promising alternative, especially as it operates in the centre of the technologically important 1260 - 1360 nm wavelength range, also known as "O band" in fiber-optic communications.

We identified molybdenum ditellurite ($MoTe_2$) as a suitable gain material for this laser. $MoTe_2$ is one of a family of two-dimensional (2D) materials that were spawned out of the graphene revolution. Many of these 2D materials, especially atomically thin transition metal dichalcogenides (TMDs) such as tungsten diselenide ($WSe_2$), molybdenum disulfide ($MoS_2$) and now $MoTe_2$, exhibit remarkable optoelectronic properties, such as strong exciton binding energy and high carrier mobility [8-12]. 2D materials are a unique group of materials whereby atoms are strongly bonded in the plane but only weakly attached out-of-plane; this weak interaction between layers makes the extraction of single or a few layers of atoms possible, which underpins this burgeoning research area [13]. Excitingly, several TMD-based lasers have been recently demonstrated, including a continuous-wave (CW) $WSe_2$ monolayer laser emitting around 740 nm at 80 K, based on a gallium phosphide photonic crystal (PhC) cavity [14], a microdisk laser including a tungsten disulfide ($WS_2$) monolayer that is sandwiched by a $Si_3N_4$ and hydrogen silsesquioxane (HSQ), operating at 612 nm at 10 K [15], and a room temperature four layer $MoS_2$ laser emitting in the wavelength range of 600 to 800 nm, based on a vertically coupled microdisk and microsphere cavity [16]. Whereas these initial demonstrations were focussed on the visible regime, a TMD-based laser operating in the silicon transparency window has now also been demonstrated [17], using a silicon nanobeam cavity and a monolayer of $MoTe_2$ as the gain material. Our work now demonstrates that the operating wavelength can be pushed further into the infrared and into the 1260 nm to 1360 nm communications window ("O band"), which signifies a technological step-change for this type of laser. We also show that a monolayer of MoTe$_2$ is not necessarily required for laser operation, by demonstrating lasing from multilayers, thereby increasing manufacturing tolerances.

**RESULTS**

**Photoluminescence from atomically thin MoTe2 flakes.** Mono- and few-layer MoTe$_2$ flakes provide the optical gain in our lasers. The layer-dependent photoluminescence (PL) of MoTe$_2$ has already been reported at both cryogenic and room temperature on a range of substrates [18,19]. At low temperature, the PL emission spectrum consists of a split peak, of which the high and low energy peaks represent the exciton and trion emission, respectively. With an increase in temperature, the energy of the trions transfers to the excitons, resulting in a redshift of the emission. In contrast to other TMDs, the PL quantum yield of MoTe$_2$ remains relatively high from mono- up to few-layers [19]. This offers some flexibility and tolerance in the preparation of 2D light-emitting samples, which make the laser fabrication easier. Additionally, we aim to operate our lasers at around 1300 nm, so relatively far from the 1130 nm emission peak of a monolayer. Since the PL emission also redshifts with an increase in layer thickness, using multilayer material is again beneficial for our application.

The atomically thin MoTe$_2$ layers studied here were exfoliated from bulk material (obtained from HQ graphene, Netherland) and then transferred onto silicon substrate by means of an all-dry transfer technique [20]. Figure 1 shows the PL emission from a 5-layer MoTe$_2$ flake on a silicon substrate at room temperature, with a peak emission at around 1157 nm.

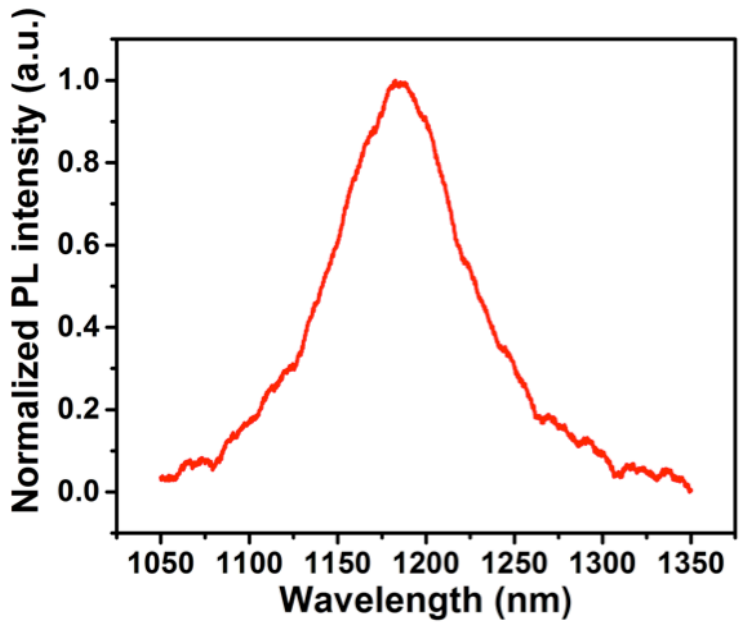

Figure 1. **Photoluminescence characterization.** Photoluminescence emission obtained at room temperature from a 5-layer MoTe$_2$ sample.

**Nanocavity for MoTe$_2$-on-silicon lasers.** At the wavelength of around 1150 nm, the intrinsic absorption loss in silicon is approximately 4 dB/cm. Although this loss is sufficiently low to enable the demonstration of reasonably high quality (Q-factor) photonic crystal (PhC) cavities [17], the practical telecommunication wavelength range starts at 1260 nm, where the loss in silicon drops well below 1 dB/cm. We therefore aimed for an operation wavelength of 1300 nm and fabricated a number of PhC nanocavities accordingly, using SOITEC silicon on insulator (SOI) material with a nominal 220 nm thick device layer on 2 μm of buried oxide.

We opted for the L3 cavity geometry, which consists of a line defect of 3 missing holes in a triangular lattice PhC [21]. Additionally, we implemented far-field optimisation in order to increase the out-of plane coupling to the cavity mode [22]. An example of such a far-field optimised PhC cavity with its corresponding radiation pattern is illustrated in Figure 2(a-b). Following fabrication (see Methods) we measure a Q-factor of Q = 5,900 for such a cavity in its "unloaded" condition, i.e. before applying the MoTe$_2$ flake onto it, with a resonant wavelength of 1298 nm. We calculate a coupling efficiency of > 90% for this cavity, using an objective with N.A. = 0.65 for the measurement (see Supplementary Information, SI). The MoTe$_2$ flakes are then precisely placed on the top of the cavity, as shown in the optical microscopic and the SEM micrographs (see Figure 2 (c) and (d)). The high-resolution AFM image indicates that the MoTe$_2$ flake on the L3 cavity is 4.0 nm thick, i.e. consisting of approximately 5 layers (more detail see SI).

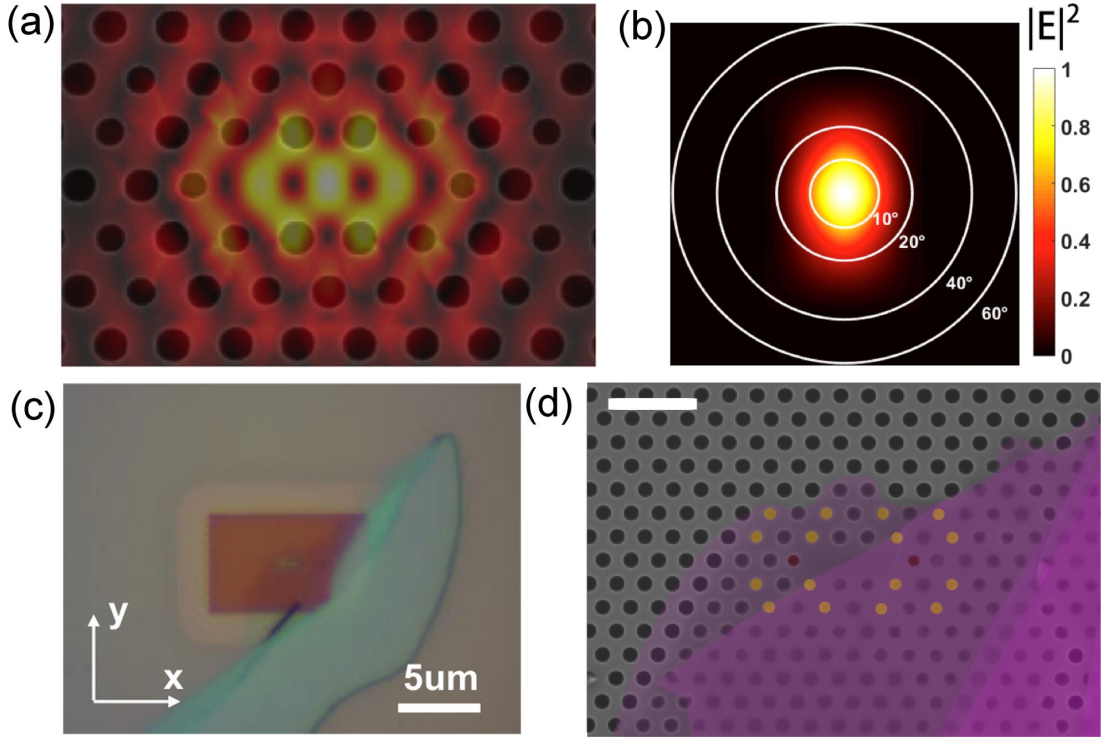

Fig. 2. **MoTe$_2$ on silicon photonic crystal cavity**. The calculated electric field in the L3 PhC nanocavity of (a) the fundamental mode M1 (b) its corresponding far-field emission (electric field intensity profile, |E|$^2$), respectively. (c) A optical microscopic and (d) a false-colored SEM image of the few-layer MoTe$_2$ on silicon PhC cavity; the scale bar in (d) represents 1 μm. We define the x and y directions in (c) to be the ΓK and ΓM direction of the PhCs, respectively.

**Dual-mode CW MoTe$_2$-on-silicon laser at room temperature.** We pump the sample with a 785 nm CW laser at room temperature and observe laser emission at 1305 nm (Mode M1 on Fig. 3). There is a second peak at 1250 nm (Mode M2), which we assign to a higher order resonance of the cavity (see SI). Here, the excitation power is measured after the objective, i.e. directly on the sample. In order to determine the emission power, we use an external 1340 nm CW laser as a calibration source and first measure the total loss of the optical elements between the sample and the CCD of the spectrometer, then calculate the actual emission power based on the photon counting number detected by the CCD.

The excitation power on the cavity for the measurement of Figure 3 is 1.9 mW, well above threshold, and the diameter of the pump spot on the silicon surface is measured to be 1.72 μm. The emission spectra were collected with a 50x objective (N.A. = 0.65) as a function of pump intensity. We note that the emission is strongly polarised perpendicular to the direction of the line defect, corresponding to laser oscillation along the defect as expected for this cavity geometry (inset of Fig. 3).

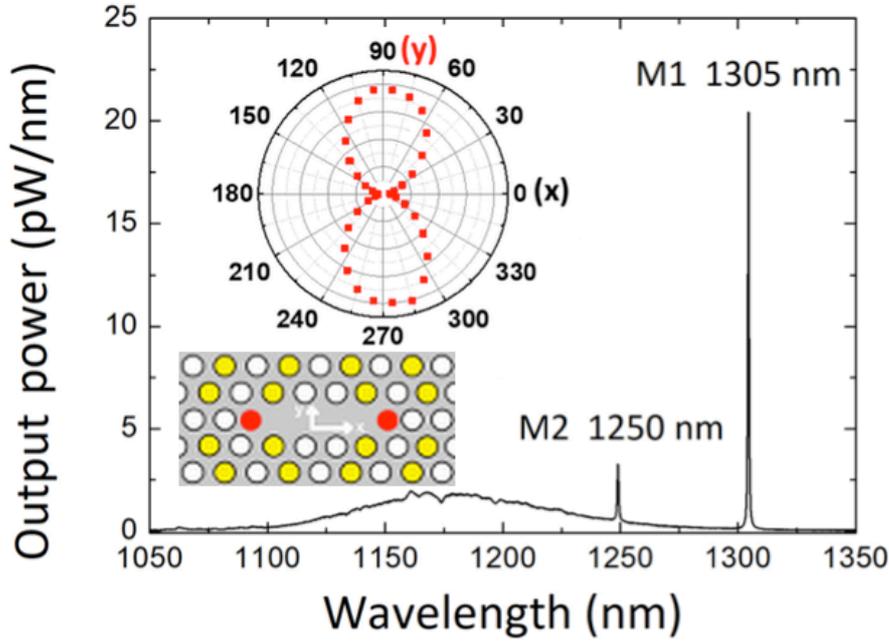

Figure 3. **Dual-mode emission and polarisation**. (a) Emission spectrum of optically pumped $MoTe_2$-on-silicon laser at room temperature with a pump power of 1.9 mW; a fundamental mode (M1) is observed at 1305 nm and a higher order mode (M2) is peaked at 1250 nm; inset: polarization characteristic of the M1 mode of the laser emission, x and y directions indicate the L3 cavity orientation (same as in Fig. 2).

The emission power, i.e. the integrated power over the fundamental cavity mode M1, as a function of pump power is plotted in Fig. 4a, showing the familiar nonlinear 'kink' around threshold. The kink is not particularly pronounced due to the very high spontaneous coupling "β" coefficient, which is discussed in more detail below. The emission power from an "off-cavity" position at the same wavelength, which does not show lasing, is also plotted in Fig. 4a for comparison. At low pump intensity, we observe an emission peak with a linewidth of 0.520 nm (full width at half maximum, FWHM), as shown in Fig. 4b. Once the pump power is increased above 35 μW, the peak narrows down to below 0.495 nm. When increasing the pump power above threshold, a linewidth re-broadening is observed, which we assign to the increase in carrier-density with pump power [17]. Fig. 4c shows the Lorentzian fits to the fundamental mode emission spectra taken at different pump powers. The "hot cavity" Q-factor calculated from these fits is Q ≈ 2,500.

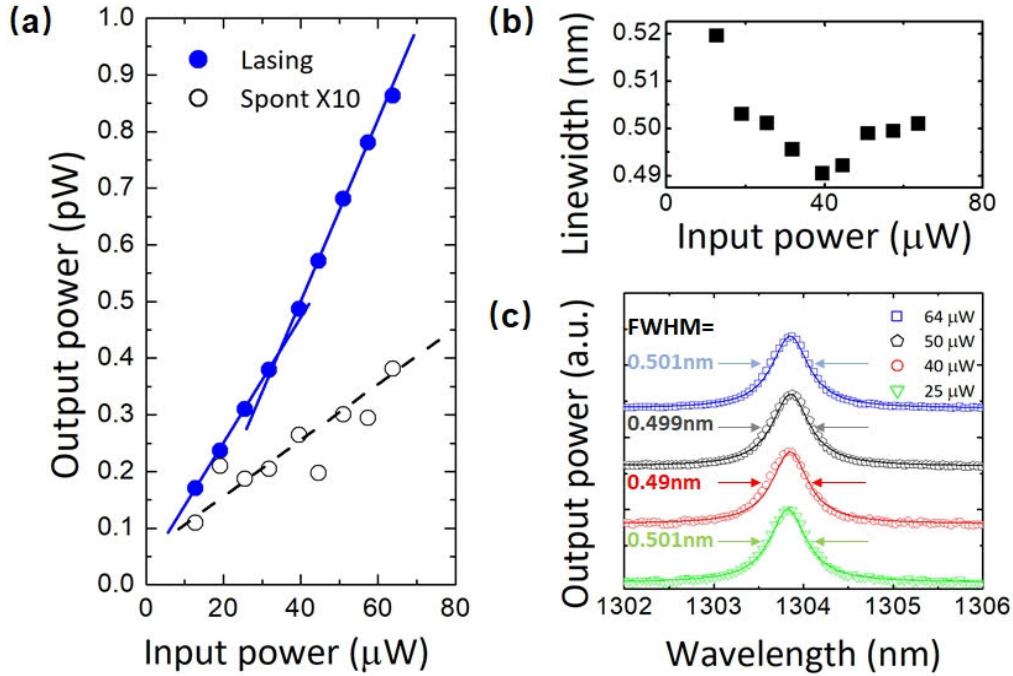

Figure 4. **Laser threshold and linewidth.** (a) The emission power of the fundamental laser mode (blue dots) and spontaneous emission at the same wavelength (black circles) as a function of pump power incident on the sample surface. The fitted straight lines are guides to the eye. The kink in the laser emission indicates a lasing threshold of 35 μW. (b) The linewidth of the emission at 1305 nm as a function of the pump power; and (c) the fundamental mode output spectra (dots) and the corresponding Lorentzian fits (solid curves) for different pump powers.

**Examining our MoTe$_2$-on-silicon laser performance.** We determine the laser emission power to be approximately 20 pW with a pump power of 1.9 mW. Taking into account that the active area of PhC cavity is 1.25 μm$^2$, the MoTe$_2$ laser emission power density is calculated to be 1.6 mW/cm$^2$. Although this output power is weaker than other compact silicon lasers [23, 24], there is still considerable scope for improvement of our MoTe$_2$-on-silicon laser, e.g. by increasing the overlap between the MoTe$_2$ gain and the cavity mode volume, leading to potential application of our 2D-on-silicon nanolasers in optical communications and interconnects.

To further verify laser operation, the β-factor, defined as the fraction of spontaneous emission coupled into the cavity mode, is obtained by fitting our experimental data, as shown in Figure 5. The best fit indicates a β-factor of 0.5 for M1 and 0.48 for M2 in our far-field optimised MoTe$_2$-on-silicon nanolaser. This suggests that in this laser, approximately 50% of the total spontaneous emission is coupled to the cavity mode. Such remarkable lasing performance is attributed to the high Q-factor of our silicon PhC cavity, which results in very strong cavity–gain coupling in the planar MoTe2-on-silicon geometry. For the fundamental mode, using the threshold value of 35 μW and the spot diameter of 1.72 μm, we determine a lasing threshold of 1.5 kW/cm$^2$.

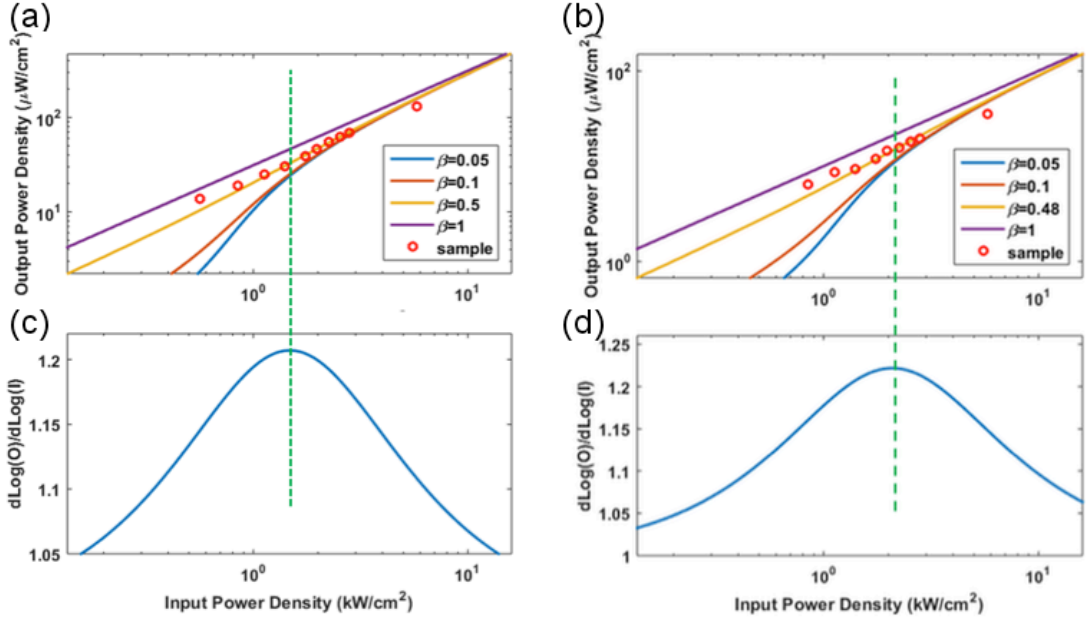

Figure 5. **Laser threshold and β-factor.** (a-b) Output power density as a function of the excitation power density of (a) M1 and (b) M2 mode. Red dots are experimental data, and the solid curve is a fit to the laser rate equation, corresponding to a spontaneous emission coupling efficiencies β = 0.5 for M1 and β = 0.48 for M2. Calculated curves for other values of β are plotted for comparison. The vertical dashed line (green) indicates the lasing threshold obtained from the fit. (c-d) First order derivative curve of the laser rate equation of M1 and M2 mode respectively - the peak position indicates the lasing threshold for each mode, for more detail see SI.

Finally, by simply changing the periods of the PhC cavity, we achieve a wide tuning range of the MoTe$_2$-on-silicon lasing wavelengths, as shown in Figure 6. The lasing emission peak can be tuned from 1127 nm to 1305 nm by varying the period of the PhCs between 272 nm and 328 nm.

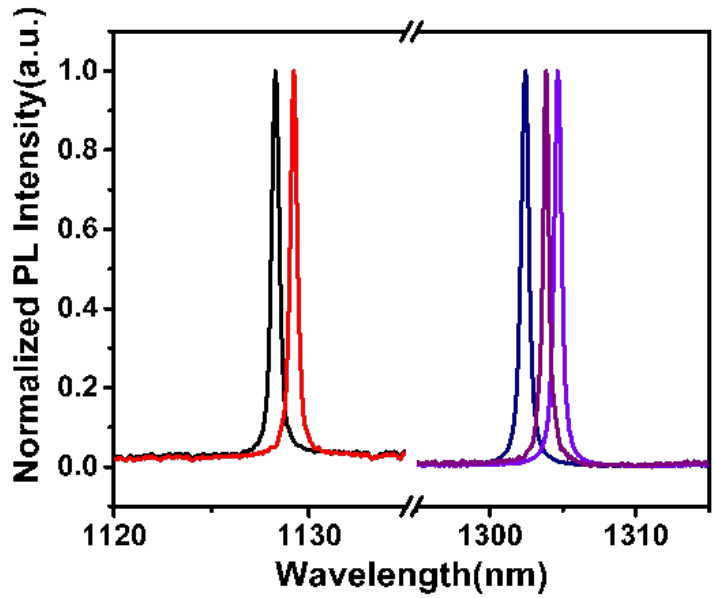

Figure 6. **Tuning of the laser emission wavelength**. The laser emission wavelength can be tuned by simply varying the period of the PhC cavities. Peaks around 1127 nm correspond to the fundamental modes from L3 cavities with periods of around 272 nm, and peaks around 1300 nm represent the fundamental modes from cavities with periods of around 328 nm.

**DISCUSSION**

It is interesting to compare our results to those already published, so to better understand the impact of the TMDC and the silicon PhC cavity, see Table 1.

Table 1. Performance comparison of various 2D material based lasers

| Reference | Gain Material | Lasing Wavelength | Operating Temperature | Pump Source | Hot cavity Q-factor* | Threshold** |
|---|---|---|---|---|---|---|
| [14] | $WSe_2$ | 740 nm | 130 K | CW | 2,466 | 1 W/cm$^2$ |
| [15] | $WS_2$ | 612 nm | 10 K | Pulse | 2,605 | 22.4 MW/cm$^2$ |
| [16] | $MoS_2$ | 600-800 nm | 300 K | CW | 3,000 | 7.14 kW/cm$^2$ |
| [17] | $MoTe_2$ | 1132 nm | 300 K | CW | 5,603 | 2.1 kW/cm$^2$*** |
| Our work | $MoTe_2$ | 1305 nm | 300 K | CW | 2,500 | 1.5 kW/cm$^2$ |

\* The Q factor is derived from lasing mode.

\*\* Calculated using the power incident on the sample surface divided by the pump spot area.

*** The threshold value is reported to be 6.6 W/cm$^2$ in [17]. Their quoted value was lower because the authors took absorption and coupling coefficients into account, whereas this was not done for the other papers.

In Table 1, we determined the threshold from the power incident on the device and the spot size as provided by the respective papers, and without taking any correction factors into account. This may lead to deviations between the values shown here and those published, e.g. for [17], but it gives a better comparison and provides more insight.

As shown in Table 1, we achieve the lowest threshold lasing at room temperature from 2D materials, although our Q factor is not the highest. We believe this can be attributed to two main reasons: 1) a suitable pump area is necessary to ensure efficient pumping of the gain materials/cavity, 2) high optical gain at room temperature is essential for low threshold operation, e.g. MoTe$_2$ is one of the best candidates so far for active materials emitting in the near-infrared.

In summary, we have demonstrated a room-temperature nanolaser based on atomically thin semiconducting material in the optical communication band for the first time. By employing molybdenum ditelluride as the gain medium, which has maximum photoluminescence efficiency at around 1150 nm (1.08 eV) at room temperature, it was surprising that we can achieve lasing at 1300 nm (0.95 eV). This demonstration thus not only contributes to the establishment of 2D TMDs as a practical novel gain medium but also has remarkable impact on silicon photonics where an on-chip silicon light sources are highly desired. With an excitation threshold as low as 1.5 kW/cm$^2$ and a reasonable emission power, our demonstration of the 2D-on-silicon nanolasers could lead to exciting applications in on-chip optical communications and interconnects. We can also foresee that with further optimisation of the cavity fabrication, i.e. fine adjustment on the PhC hole sizes to achieve even higher Q cavities and employ L5, L7 cavity geometries to enhance the far-field emission, we will be able to further reduce the laser threshold and increase the laser output power.

**Methods**

**Design and characterization of the PhCs cavity**
To provide the feedback for the active gain material MoTe2, we used a far-field optimised L3-type planar PhC cavity structure. We performed 3D FDTD simulation to obtain the desired cavity modes. Firstly, we fixed the value of *r/a*, the ratio of the radius of the PhCs (*r*) and the period (*a*), and then changed *r* and *a* respectively to obtain the resonant wavelength of the PhC cavities to be around 1300 nm. Secondly, the Q-factors and far-filed were optimized by changing the PhC hole size (*r*) and positions around the L3 cavity. The optimized parameters of the final design are included in SI. This design achieved a good compromise between Q-factor and the far-field emission pattern of the M1 mode. The Q-factor of the PhC cavity is measured by detecting the cross-polarised resonantly scattered light from a normally incident white light source.

**Fabrication of the PhC cavity**

The cavity was fabricated on a SOITEC SOI wafer comprising a 220 nm thick silicon layer on a 2 μm sacrificial silicon dioxide layer. The pattern was firstly defined in ZEP520A electron beam resist by electron-beam lithography (Raith Vistec EBPG5000+ 100kV), and then transferred into the silicon membrane using an inductively coupled plasma system (Oxford PlasmaPro 100ICP180) with HBr gas. The residual resist was removed by a microresist remover 1165. The silica beneath the PhC cavity was finally removed with the sample submerged in a hydrofluoric acid bath.

**Exfoliation and transferring the 2D flakes (Hybrid device fabrication)**

To prepare 2D materials, the molybdenum ditelluride flakes were initially exfoliated onto a PDMS stamp. Their thicknesses were confirmed using an optical microscope and an atomic force microscope. After preparing the 2D material with required layer number, we used the typical dry transfer technique [2D Materials 2014, 1(1): 011002.] to transfer the few-layer MoTe2 on to the silicon PhC cavity.

**Laser performance characterisation**

Pump power density is an important parameter of any laser performance. Therefore, we need to measure the size of the pump beam precisely. Firstly, we used a gold marker in 10 μm size to calibrate our microscope imaging system with a CCD, acquiring that the imaging size of each pixel is 0.187 μm. Then the pump spot on the silicon sample surface was captured by the CCD camera. The spot intensity data was acquired by the grey value profile and then fitted with a Gaussian curve to obtain the FWHM. The FWHM of the pump beam in our laser measurements was measured to be 1.72 um, as shown in SI.

To optically excite the MoTe2-on-silicon laser, a 785 nm CW laser was focused through a 50x objective (NA=0.65) onto the active gain material. The spectra were collected by an IR spectrometer with a 600 line/mm grating (Princeton Instrument SP2758), which corresponds to a spectral resolution of 0.2 nm, and a nitrogen cooled CCD (Princeton Instrument OMA-V:1024/LN).

**Acknowledgements**

This work is supported by National Key R&D Program of China (2016YFA0301300), National Natural Science Foundation of China (11674402, 11334015), Guangzhou science and technology projects (201607010044, 201607020023), Natural Science Foundation of Guangdong (2016A030312012), the Fundamental Research Funds for the Central Universities, and the EPSRC Grant EP/M015165/1 (Ultrafast Laser Plasma Implantation-Seamless Integration of Functional Materials for Advanced Photonics). We would also like to acknowledge Prof. Xuetao Gan at Northwestern Polytechnical University, China for useful discussions on exfoliation and transferring the 2D flakes.